\documentstyle{mn}
\newcommand{\rmn}[1] {\mathrm{#1}}

\title[Partition function based analysis of CMB maps]
      {Partition function based analysis of CMB maps}
 
\author[J.M Diego et al.] 
   { J.M  Diego,$^{1,2}$, E. Mart\'\i nez-Gonz\'alez,$^1$, J. L.
Sanz,$^1$
\newauthor Silvia Mollerach,$^{3,4}$ \& Vicent J. Mart\'\i nez$^3$ \\
   $^1$Instituto de F\'\i sica de Cantabria, Consejo Superior de
   Investigaciones Cient\'\i ficas Universidad de Cantabria,
   Santander, Spain\\
   $^2$Departamento de F\'\i sica Moderna, Universidad de Cantabria,
   Avda. Los Castros s/n, 39005 Santander, Spain\\
   $^3$Departamento de Astronom\'\i a y Astrof\'\i sica, 
Universitat de  Val\`encia, 46100 Burjassot, Valencia, Spain\\
   $^4$Departamento de F\'\i sica, Universidad Nacional de La Plata,
   c.c. 67, 1900 La Plata, Argentina}

\pagerange{\pageref{firstpage}--\pageref{lastpage}}

\begin{document}

\maketitle

\label{firstpage}

\begin{abstract}
We present an alternative method to analyse cosmic microwave
background (CMB) maps.
We base our analysis on the study of the partition function. 
This function is used to examine the CMB maps making use of the different
information embedded at different scales and moments. \\
Using the partition function  in a likelihood analysis
in two dimensions ($Q_{{\rmn rms}-{\rmn PS}}$,$n$), we find the 
best-fitting model
to the best data available at present (the {\it COBE}--DMR 4 years data set).
By means of this analysis we find a maximum in the likelihood
function for $n=1.8_{-0.65}^{+0.35}$ 
and $Q_{{\rmn rms}-{\rmn PS}} = 10_{-2.5}^{+3} \ \mu$K (95 \% 
confidence level) in agreement with the results of other similar 
analyses (Smoot et
al. 1994 (1 yr), Bennet et al. 1996 (4 yr)).\\
Also making use of the partition function we perform a multifractal
analysis and study the possible fractal nature of the CMB sky.
We find that the measure used in the analysis is not a fractal. \\
Finally, we use the partition function for testing the statistical 
distribution of the {\it COBE}--DMR data set. We conclude that no evidence of
non-Gaussianity can be found by means of this method.

\end{abstract}


\section{Introduction}

In a few years the forthcoming CMB data sets from the missions MAP
(NASA)
and PLANCK (ESA) will offer us a much better image of
the young universe than ever before. 
The CMB represents a view of the universe when it
was about $0.002\%$ of its present age. 
CMB anisotropies provide a link between theoretical predictions and 
observational data. Undoubtedly these data will constrain more accurately the fundamental
cosmological parameters.
In recent years several groups have been very active
in the study of the CMB anisotropies.
Many statistical methods have been adapted to the analysis of the
future CMB maps, and  others are being developed.\\
There are many methods which can give relatively accurate values
for the parameters of the cosmological models. For example, the
power spectrum is considered to be
the best discriminator between different models  
(Bond et al. 1997, Hinshaw et al. 1996a, Wright et al. 1996, Tegmark  1996). 
Related through a 
Legendre expansion
to the power spectrum, the two-point correlation
function is also a useful discriminator
(Cay\'on et al 1996, Hinshaw et al 1996b).
These analyses, based on the power spectrum, are considered as
{\it classical} but there are many other methods that do not make
use of the power spectrum.
The quality of the CMB maps demands for other statistics to
supplement the power spectrum, looking for instance at 
 morphological or topological
characteristics of the data. For example the one-dimensional analysis  
is a geometrical method useful for
one-dimensional scans of CMB data and is based on the study of 
regions above or below a certain level (Guti\'errez et al. 1994). 
The peak analyses, similar to the previous 
one but for two-dimensional data, deal with the number of spots above a
given threshold (Fabbri \& Torres 1995) or with other geometrical 
properties like the Gaussian curvature or excentritity of the maxima 
(Barreiro et al. 1997). Other approaches are the genus (Smoot et al. 1994, 
Torres et al. 1995), Minkowski functionals, which relate several 
geometrical aspects at the same time (Schmalzing \& G\'orski 1997, Winitzki \&
Kosowsky 1997), wavelet based techniques (Pando et al. 1998, Hobson et al. 
1998) who have shown wavelets to be very effective at detecting 
non-Gaussianity in the CMB, and fractality (Pompilio et al. 1995, 
De Gouveia dal Pino et al. 1995, Mollerach et al. 1998).\\

In this paper we present an alternative method to analyze CMB maps based 
on the partition function. 
This function contains useful information about the temperature 
anisotropies at the different scales and moments.
The method presented here is related  to the one used by Smoot et al. (1994)
based on moments at different smoothing angles. However, our method is more
general and powerful because it works with any moment, not
only with positive and integer ones.\\
The structure of the paper is as follows.
In Section 2 we present the partition function and discuss
its main characteristics. In the same section
three different analyses based on that function
are introduced: a likelihood analysis, a multifractal analysis and a
test of Gaussianity.
The likelihood analysis uses the partition function to search
for the parameters ($Q_{{\rmn rms}-{\rmn PS}}$ and $n$) that best
fit a given data set. The multifractal analysis searches for
scaling laws and fractal behavior of the data.
There are theoretical reasons (Sachs-Wolfe effect) to expect
scaling in the data. Here we present the generalized fractal
dimensions and the scaling exponents and also comment  on the
possible multifractality of the CMB sky.
In a recent paper, Ferreira et al. (1998)
found evidence of non-Gaussianity in {\it COBE}--DMR data at $99 \%$
confidence level. We show that the partition function  can
be used to study this property. There is a clear relation
between the partition function  and the cumulant function,
the last one having a specific form for a Gaussian signal.
In section 3 we apply the results of the previous sections to the 
{\it COBE}--DMR 4 years data set and we compare with other results. 
We conclude in section 4.


\section{The partition function  }
Let us start directly with the definition,
\begin{equation}
 Z(q,\delta) =\sum_{i=1}^{N_{\rmn boxes}(\delta)} \mu_i(\delta)^q
\end{equation}
where $Z(q,\delta)$ is the partition function. The quantity
$\mu_i(\delta)$ is called the {\it measure}, it is a function
of $\delta$ which is the size or scale of the boxes used to cover the 
sample. The boxes are labeled by $i$ and  $N_{\rmn boxes}(\delta)$ 
is the number of
boxes (or cells) needed to cover the map when the grid with
resolution $\delta$ is used. The exponent $q$ is a continuous 
real parameter that plays the role of the order of the moment of
the measure.\\ 
Let us consider a map of $N$ pixels. Now the map is divided
in boxes of size $\delta\times\delta$ pixels and the measure
$\mu_i(\delta)$ is computed in each one of the resulting boxes.
Changing both, $q$ and $\delta$, one calculates the function
$Z(q,\delta)$. We would like to emphasize that the calculation of 
$Z(q,\delta)$ is $O(N)$. \\
One is free to make any choice of the measure $\mu(\delta)$ 
provided that  several conditions are satisfied, 
the most restrictive  being $\mu_i(\delta) \ge 0$.  
There are no general rules to decide
which is the best choice. For CMB maps, we use the most natural
measure defined as follows :
\begin{equation}
 \mu_i(\delta)=\frac{1}{T_*} \sum_{{\rmn pix}_j \in {\rmn box}_i}
 T_{{\rmn pix}_j}.
\end{equation} 
Thus the measure in the box $i$ is the sum of the {\it absolute}
temperatures $T_{\rmn pix}$ of the pixels  
inside the box in units of Kelvin. This is a very natural measure
comparable to the measure used in the study of galaxies or clusters
distribution (Mart\'\i nez et al. 1990, see Borgani 1995 for a review),
where the measure is taken as the
total mass (or the total number of galaxies/clusters) contained
in the box. The constant $T_*$ 
is a normalization constant. 
The measures are interpreted as probabilities and they have to be 
normalized, i.e $\sum_i \mu_i = 1$. So $T_*$ is simply the sum of 
the absolute temperatures over all pixels and therefore is a constant 
for all boxes and scales.\\

The temperature in the pixels is almost the same everywhere because of 
the homogeneity of the signal, and one expects that different models will 
behave in a very similar way, making difficult the task of distinguishing them.
We shall show how the partition function  overcomes this problem. \\
Alternatively, Pompilio et al. (1995), in a multifractal analysis of string 
induced CMB anisotropies (one dimensional scans) used as measure:
\begin{equation}
  \mu_i(\delta) = \sum_{j=i-M\delta /2}^{j=i+M\delta /2}
\lbrack \Delta_j - \Delta_{j+1} \rbrack^2,
\end{equation}
where $\Delta_j$ denotes the fluctuation of the temperature in pixel $j$
with respect to the mean. 
Here M is the total number of points in the data set, and
$(i-M\delta /2)$ and $(i+M\delta /2)$ are the lower and
upper edges of the $i$th segment with $M\times\delta$ points,
centered on the $i$th point of the scan. The scale $\delta$
runs between $1/M$ for the smallest segment and 1 for the
whole segment.
However, this measure is not sensitive to the sign of the
temperature fluctuations because of the square in its definition. 
Due to this fact the full information of the fluctuations is not
conveniently considered. In addition, the generalization
 of this measure to 
2D maps is not unique. \\
  
Using the measure proposed in this paper, the differences between 
two temperature
data sets appear when high values of the exponent $q$ are considered.
The method is able to
differentiate between two very close models with $q$ ranging 
between  $[-2.5\times10^5, +2.5\times10^5]$. This range for
$q$ is in agreement with the level of inhomogeneity. 
We are using {\it absolute} temperatures, that is, we have inhomogeneities
of order $10^{-5}$ with respect to the mean value and the signal
is almost flat. One can consider $q$ as a powerful microscope,
able to enhance the smallest differences of two very similar
maps. Furthermore, $q$ is a  selective parameter. Choosing large values
of $q$ in the partition function, favors contributions from cells
with relatively high values of $\mu_i(\delta)$ since $\mu_i^q \gg
\mu_j^q$ for $\mu_i > \mu_j$, if $q \gg 0$. Conversely, $q \ll 0$
favors the cells with relatively low values of the measure. This
is the role played by the moments, changing $q$ one explores 
the different parts of the measure probability distribution.   
The other parameter, $\delta$, acts like a filter. Choosing big values of
$\delta$ is similar to apply a large scale filter to the map. One looks
at different scales when the parameter $\delta$ is changed.\\

To summarize, $Z(q,\delta)$ contains information at different scales
and moments. The multi-scale information
gives an idea of the correlations in the map, meanwhile the moments
are sensitive to possible asymmetries in the data, as well as some
deviations from Gaussianity. 
In what follows we show the power of the partition function to
extract
useful information from CMB data. Three different analyses 
are used for
 this purpose.

\subsection{Likelihood analysis}

We shall use the partition function to encode the information of
a given map. We compute it both for the experimental data and for
simulated ones corresponding to different models. 
In this process we are comparing the data
and the models at several scales and using different moments. If
there are some differences at some scale or moment, then the
partition function  should make it evident. 
The likelihood function will have a maximum for the best-fitting 
model to the data. For the CMB maps analyses, we
consider models corresponding to different values of the
spectral index $n$ and the normalization 
$Q_{{\rmn rms}-{\rmn PS}}$.\\

The likelihood is defined in the usual way (assuming a Gaussian 
distribution for $\ln Z(q,\delta)$). We work with ${\cal Z} = 
\ln Z(q,\delta)$ 
instead of $Z(q,\delta)$ because of the large values of $q$ which make 
imposible to compute directly $Z(q,\delta)$,
\begin{equation}
 L(Q_{{\rmn rms}-{\rmn PS}},n) = \frac{1}{(2\pi)^{n/2}(detM)^{1/2}}
\exp(-\frac{1}{2}\chi^2),
\end{equation}
where,
\begin{equation}
 \chi^2=\sum_{i=1}^{N_{p}}\sum_{j=1}^{N_{p}}(\langle {\cal Z}(i) \rangle -
{\cal Z}_D(i))M_{ij}^{-1} (\langle {\cal Z}(j) \rangle - {\cal Z}_D(j)),
\end{equation}
and $\langle {\cal Z}(i) \rangle$ is the average of the  ${\cal Z}$
for the $N_{{\rmn rea}}$ realizations of the model at bin $i$. The
index $i$ defines pairs of values ($q$,$\delta$) and runs from 1  
to $N_q \times N_{\delta}$. That is, $i$ runs from 1 to the total number
of points $N_p$ where $Z(q,\delta)$ is defined.\\
${\cal Z}_D(i)$ is the value of ${\cal Z}$ for the experimental data
at bin $i$. $M_{ij}$ is the covariance matrix calculated with 
Monte Carlo realizations:
\begin{equation}
 M_{ij}=\frac{1}{N_{{\rmn rea}}}\sum_{k=1}^{N_{{\rmn
rea}}}({\cal Z}_k(i)-\langle
{\cal Z}(i) \rangle)({\cal Z}_k(j)-\langle {\cal Z}(j) \rangle).
\end{equation}
${\cal Z}_k(i)$ denotes the value of ${\cal Z}$ at bin $i$ for the
$k$ realization.\\
We tried different number of realizations $N_{{\rmn rea}}$ but the
results appear to be stable for $N_{{\rmn rea}} > 2000$
per value of $Q_{{\rmn rms}-{\rmn PS}}$ and $n$.\\
We have two possibilities to perform a best fit to the data. The first
one is to minimize $\chi^2$ and take the values of
$Q_{{\rmn rms}-{\rmn PS}}$ and $n$ at the  minimum of the $\chi^2$
surface as the best-fitting values.
The second possibility is to work with the likelihood $L$ looking
for the maximum. We tested the two possibilities using
simulated CMB maps derived from a given pair of parameters
($Q_{{\rmn rms}-{\rmn PS}}$, $n$) and then using these maps as the data
maps.
Due to cosmic variance we obtain  a set of maxima in the
likelihood and of minima in the $\chi^2$. The conclusion is
that the likelihood is somewhat better than the $\chi^2$ as
expected. For instance with 2000 input realizations with
$Q_{{\rmn rms}-{\rmn PS}}=14 \  \mu$K and $n=1.3$ the distribution of 
maximun likelihood values finds a
maximum at $Q_{{\rmn rms}-{\rmn PS}}=13_{-4.25}^{+3} \  \mu$K and 
$n=1.2_{-0.15}^{+0.8}$ while the $\chi^2$ renders a minimum in 
$Q_{{\rmn rms}-{\rmn PS}}=17_{-4}^{+5} \  \mu$K and
$n=1.0_{-0.35}^{+0.45}$.
The errors are marginalised at the 68\% confidence level and are similar
to those obtained with the standard methods based on the power 
spectrum (see for instance Wright et al. 1996).

\subsection{ Multifractal analysis}

The notion of multifractal measure was first introduced
by Mandelbrot (Mandelbrot 1974) in order to study different
aspects of the intermittency of turbulence  (see also Sreenivasan
and Menevau 1988).
The multifractal formalism was further developed by many other
authors and today it is a standard tool applied in almost all
 fields of science: 
molecular physics, biology, geology, astronomy, etc .
In the context of the description of the large scale structure
of the Universe it was first introduced by Jones et al. (1988).

Some authors (Pietronero \& Sylos Labini 1997) suggest that the
distribution of matter in the universe is fractal
with dimensionality $D_2 \simeq 2$. They
defend that the scaling remains up to the larger
scales probed by the present day available redshift catalogues.
Many other authors, however, have found enough evidence of homogeneity
at large scales (Davis 1997, Guzzo 1997, Scaramella et al. 1998)
in the analysis of the same data sets.
One of the basic tenets of the standard cosmology is that
at very large scales the distribution
of matter is homogeneous. The homogeneity  and isotropy
of the CMB support this overwhelming evidence, indicating that
there exists a continuous transition between scale invariant
clustering at small scales and homogeneity at large scales
(Mart\'\i nez et al. 1998; Wu, Lahav \& Rees 1998).\\

At large angular scales, the CMB anisotropies $\Delta T/T$ generated 
from a scale-free density perturbation power spectrum in a flat
$\Omega = 1$ universe can be described by a fractional Brownian 
fractal (as shown in Mollerach et al. 1998). 
In particular, both inflationary and defect models predict an
approximately scale invariant {\it Harrison-Zel'dovich}
spectrum on large angular scales showing the scaling predicted 
by the Sachs-Wolfe effect. At small angular scales
$(0.2^{\circ} \leq \theta \leq 1^{\circ})$ the predictions of inflation
and topological defects models are different (Durrer et al 1997)
allowing to differentiate them.
It is then interesting to study the possible fractality of the CMB
anisotropies, since the seeds or fluctuations that are supposed to 
be the precursors of the largest structures observed today, are yet
unperturbed by evolutionary phenomena. Several works follow
this kind of analysis. In the paper by De Gouveia dal Pino et al.
(1995), the authors based their analysis in the study of the
perimeter-area relation of the isocontours of temperature
at a given threshold. They used the {\it COBE}--DMR 1 year data
set and only the 53 GHz channel. They
found evidence for a fractal structure in the {\it COBE}--DMR data with
dimension $D = 1.43$ suggesting that the CMB could not be homogeneous.
Apart from the fact that these data have a low signal to noise
ratio, this does not necessarily mean that the CMB is not
homogeneous. This dimension corresponds to 
the temperature isocontours, and not to the temperature itself.
Other works use multifractal analysis with CMB. Pompilio et
al. (1995) apply the multifractal analysis to simulated string-induced
CMB scans searching for the non-Gaussian behavior induced by
cosmic strings. More recently, Mollerach et al. (1998) have
applied a fractal analysis in order to study the roughness of the
last scattering surface and used this technique to search for the
model
that best fit the {\it COBE}--DMR 4yr data. These authors show the
capabilities of this method for the analysis of future data, in particular
for those experiments with high signal to noise ratio.\\
        
In this section we will use the partition function
to study the possible multifractality of the CMB sky, 
using as measure the absolute temperature (see eq. 2).
The multifractal analysis has been presented in several
versions but the most popular is due to  Frisch and Parisi
(1985), Jensen et al. (1985) and Halsey et al. (1996), where the
spectrum of singularities $f(\alpha)$ was introduced.
We will give here a brief description of the multifractal approach.
A presentation of the method can be found in Feder 1988, 
Schuster 1989, Vicsek 1989 and more formally in Falconer 1990.\\  
The multifractal formalism has as starting point 
the partition function. The generalized or Renyi dimensions
are defined by the asymptotic behavior (as the scale $\delta$
tends to zero) of the ratio between $\ln Z(q,\delta)$ and
$\ln \delta$,
\begin{equation}
 D(q) =  \lim_{\delta \to 0} \frac{1}{q-1}
\frac{\ln Z(q,\delta)}{\ln \delta}. \label{renyi}
\end{equation}
It is easy to see that for $q=0$ we obtain the box-counting or
capacity dimension,
\begin{equation}
 D(0) =  \lim_{\delta \to 0}
\frac{\ln {N_{\rmn boxes}(\delta)}}{\ln (1/\delta)}. \label{boxdim}
\end{equation}
For $q=1$, $D(1)$ is the information dimension, which
is obtained from Eq. (7) 
by applying L'H\^opital's
rule. For $q=2$, $D(2)$ is the correlation dimension
(see Schuster 1998 for other alternative definitions and the
relation between them).
A {\it simple} fractal or {\it monofractal} is defined by a 
constant $D(q)$. Dependence of $D$ on $q$ defines a multifractal. 
In most of the practical applications of the multifractal
analysis, the limit in Eq. (7) 
cannot be calculated, 
either because we do not have information for small distances 
(as it happens in this case) or because below a
minimum physical length no scaling can exist at all (for example
the size of a galaxy in the multifractal nature of the
galaxy distribution).
This problem is usually overcome by finding a scaling range
$[\delta_1, \delta_2]$ where a power--law can be fitted to the
behavior of the partition function
\begin{equation}
Z(q,\delta) \propto \delta^{\tau(q)} \quad\quad \mbox{for} \quad
\delta_1 \le \delta \le \delta_2. \label{scaling}
\end{equation}
The scaling exponents $\tau(q)$ are related with the generalized
dimensions by
\begin{equation}
\tau(q) = (q-1) D(q). \label{tqdq}
\end{equation}
Other quantity, commonly used in the characterization of
multifractals, is the so--called $f(\alpha)$ spectrum.
If for a given box (labeled by $j$) the measure scales as
\begin{equation}
 \mu_j(\delta) \sim \delta^{\alpha_j} ,
\end{equation}
then, the exponent $\alpha$, which depends in principle on the position is
known as crowding index or H\"older exponent.
If all the points have the same scaling, then all the exponents
$\alpha$ will be the same and this corresponds to a monofractal. 
Otherwise, if we have boxes with
different scaling, what we have is a mixture of monofractals. 
This set is known as a {\it multifractal} (each monofractal
formed by the points with the same scaling and therefore with
the same exponent $\alpha$). The exponent $\alpha$ is used to
label the boxes covering the set supporting a measure,
thereby allowing a separate counting for each value of $\alpha$.
In a multifractal set
$\alpha$ can take different values within a certain
range, corresponding to the different strength of the measure
(Halsey et al. 1996).
The subset formed by the boxes with the same $\alpha$ will be
denoted $S_{\alpha}$. 
This subset has $N_{\alpha}(\delta)$ elements (boxes)
and in general, for a multifractal set,
this number varies with the scale $\delta$ as
\begin{equation}
 N_{\alpha}(\delta) \sim \delta^{-f(\alpha)} .
\end{equation}
Comparing this expression with the definition of the
box-counting dimension, Eq. (8), 
the quantity $f(\alpha)$ can be interpreted as the fractal dimension 
of the subset  $S_{\alpha}$. However, this physical meaning
of the function $f(\alpha)$ is not always true (Grassberger,
Badii \& Politi 1988; Falconer 1990).\\

It can be shown (Halsey et al. 1996; Mart\'{\i}nez et al. 1990)  
that the quantities $q$ and $\tau (q)$ can be related through a
Legendre transformation with $\alpha$ and $f(\alpha)$.
These relations are:
\begin{equation}
 \alpha (q) =  \frac{d\tau (q)}{dq}, \label{leg1}
\end{equation}
\begin{equation}
 f(\alpha) = q\alpha (q) - \tau (q). \label{leg2}
\end{equation}
To illustrate this section we use
the well known multiplicative multifractal cascade
(Meakin 1987; Mart\'\i nez et al. 1990). The construction
of this multifractal is as follows:
A square is divided into four equal square pieces and a
probability $p_i$, $(i=1,\dots,4)$, such that $\sum_{i=1}^4 p_i=1$,
is assigned to each one.
Each piece is again subdivided in four small squares, allocating
again a value $p_i$ randomly permuted to each one. The measure
assigned to each one of the new subsquares is the product
of this value of $p_i$ and the corresponding value of its parent
square. The subdivision process is continued recursively.
In Fig. 1 we show a realization of this multifractal on a
grid of $256 \times 256$ pixels for the values of the probabilities
$p_1=0.18$, $p_2=0.23$, $p_3=0.28$ and $p_4=0.31$.
We can easily calculate the theoretical values of the multifractal
functions $D(q)$ and $f(\alpha)$  for this illustrative example
(Mart\'\i nez et al. 1990).
With the multiplicative multifractal we tested the power of
the method to recover the true dimensions.
In Fig. 2 we show the generalized
dimensions $D(q)$ and the corresponding spectrum
of fractal dimensions $f(\alpha)$. These curves
match perfectly the theoretically expected ones.
Note that a single monofractal should
render a straight line for $D(q)$ and a single point for $f(\alpha)$.\\
 
\subsection{Testing Gaussianity}

A Gaussian distribution of CMB temperature fluctuations is a
generic prediction of inflation. Forthcoming high-resolution maps
of the CMB will allow detailed tests of Gaussianity down to
small angular scales, providing a crucial test of inflation.
Most of the works that analyse CMB maps assume Gaussian initial
fluctuations. 
Kogut et al. (1996) find that the genus,
three-point correlation function, and two-point correlation
function of temperature maxima and minima are all in good 
agreement with the hypothesis that the CMB anisotropy on angular
scales larger than $7^{\circ}$ represents a random-phase Gaussian field.
Other alternative methods are proposed, like the angular-Fourier
transform  (Lewin et al. 1998), Minkowsky functionals (Schmalzing
\& G\'orsky 1998), correlation of excursion sets (Barreiro et al. 1998),  
and the bispectrum (Heavens 1998). In an 
analysis of the 4 years {\it COBE}--DMR data based on the bispectrum 
Ferreira et al. (1998) have found that Gaussianity is ruled out 
at a confidence level in excess of $99 \%$ near the multipole of 
order $l=16$.\\

In this section we will test the Gaussianity of the CMB data 
using an alternative method. The idea is to use
the relation between the partition function  and the generating
function, the last one defined as,
\begin{equation}
 G_x(t) = \langle e^{tx} \rangle .
\end{equation}
If we know that $x$ is Gaussian distributed then,  solving the
integral corresponding to the mean value of the previous definition,
results :
\begin{equation}
 G_x^{\rmn Gauss}(t) = e^{t \langle x \rangle + \frac{t^2\sigma_x^2}{2}}.
\end{equation}
It follows from the last expression that the cumulant function   $F$ is:
\begin{equation}
 F(t) = \ln G(t) = t \langle x \rangle + \frac{t^2\sigma_x^2}{2} .
\end{equation}
Finally, the function
\begin{equation}
   H(t) = F(t) - t \langle x \rangle - \frac{t^2\sigma_x^2}{2} 
\end{equation}
should be zero for all $t$ for a Gaussian field. \\
Let us consider the definition of $Z(q,\delta)$. If the measure is
defined as 
\begin{equation}
  \mu_i^*(\delta)=e^{\mu_i(\delta)},
\end{equation}
with $\mu_i(\delta)$ the same as in section 2. 
Then the partition function is
\begin{equation}
  Z(q,\delta) = \sum_{i=1}^{N_{\rmn boxes}(\delta)} e^{q\mu_i(\delta)},
\end{equation}
or equivalently,
\begin{equation}
 Z(q,\delta) = N_{\rmn boxes}(\delta)\langle e^{q\mu} \rangle =
N_{\rmn boxes}(\delta)G_{\mu}(q).
\end{equation}
This relation between $Z(q,\delta)$ and $G_{\mu}(q)$ allows us to
construct the function $H(q)$ which,
for a Gaussian measure $\mu$, 
 should be zero for all $q$ at
each scale $\delta$.
This is a simple way to find non-Gaussian signals. The function
$H(q)$ represents the contribution of {\it all} the moments larger
than 2. This contribution should be zero only for a Gaussian field.
A plot of this function indicates directly the deviations from
Gaussianity.
 
\section{Results: Application to {\it COBE}--DMR data}

As a practical use of the methods presented we will apply them 
to the 4 years {\it COBE}--DMR data.

\subsection{Description of the data}

We use the {\it COBE}--DMR 4 years $53 + 90$ GHz maps combination,
which is the choice with the largest signal to noise ratio
(Bennet et al. 1996). These data are in the {\it Quad-Cube}
pixelization with a pixel size of $\sim 2.6^{\circ}$ and the
resulting number of pixels is 6144. The data in each pixel  
represents $\Delta T$ in $mK$ units. The dipole has already been
subtracted. Assigned to each pixel there is an additional information,
 the number of times that this
part of the sky was explored by the antenna. This information
is relevant for the estimation of the instrumental noise.\\
Part of the data is contaminated by the galactic emission. There
is a strip between $ \sim \pm 20^{\circ}$ (in galactic coordinates)
in which the galactic emission dominates the CMB signal. This
strip should not be included in the analysis in order to avoid spurious
signals. In addition to this strip there are two patches in the
sky (one near Orion and the other one in Ophiucus)  that show a
strong galactic emission at mm wavelengths (Cay\'on et al. 1995), 
and should therefore also be removed from the analysis.
When this mask is applied, the number of surviving pixels reduces to  3881 
from the original 6144.

\subsection{Likelihood analysis}
In order to determine which are the values of the quadrupole normalization 
$Q_{{\rmn rms}-{\rmn PS}}$ and the spectral index $n$ that best fit
the {\it COBE} data,
we perform Monte Carlo simulations of the CMB maps for a scale-free
model with a power
spectrum given by $P(k) \propto k^n$, which has variance in the
$a_{lm}$ multipoles given by (Bond and Efstathiou 1987):
\begin{equation}
 C_l = \frac{4\pi}{5}Q_{{\rmn rms}-{\rmn PS}}^2\frac{\Gamma[l + (n -
1)/2]
\Gamma[(9 - n)/2]}{\Gamma[l + (5 - n)/2]\Gamma[(3 + n)/2]}.
\end{equation}
We consider different values for  $Q_{{\rmn rms}-{\rmn PS}}$ and the $n$
ranging from $Q_{{\rmn rms}-{\rmn PS}} = 4  \mu$K to $Q_{{\rmn
rms}-{\rmn PS}} = 35 \mu$K 
and from $n=0.3$ to $n=2.3$. 
We add instrumental noise based on the number 
of data collected by {\it COBE}--DMR at each pixel.
Furthermore, there is another effect that must be taken into account,
the cosmic variance. To treat conveniently this effect we
perform a large number of simulations ($\geq$ 2000) for
each pair of values ($Q_{{\rmn rms}-{\rmn PS}}$, $n$) and then 
we compare the average ${\cal Z} = \ln Z(q,\delta)$ values of 
these simulations with the ${\cal Z}$ corresponding to the {\it COBE}--DMR 
data (the used values for $q$ and $\delta$ were $q=-120000,-40000,72000,
152000$ and $\delta=3,4,8,16$ pixels).
The choice for $q$ and $\delta$ values is based on the test described in the
last part of section 2.1. 
The size of the $Z(q,\delta)$ grid, $N_q \times N_{\delta}$ is not critical 
 and what is now relevant is the $q$ values
considered. In particular, high order moments (i.e large $q$) are very
sensitive to the tail of the distribution and therefore the results obtained
with those high values on the parameter estimates are not stable.
The combination of $q$ and $\delta$ values, 
was one of the combinations for wich the recovered parameters $Q_{\rmn rms}$ 
and $n$ were closer to the input parameters and with smaller error bars. 
As mentioned in section 2,
 $q$ should take values of order $10^5$ in order to distinguish between models
 with temperature fluctuations of order $10^{-5}$. The values of $q$ where
chosen to be asymmetric in an attempt to consider possible asymmetries that
could exist between the negative and positive temperature
 fluctuations. The range for $\delta$ runs from 2 pixels (approximately the
 antenna size) to 24 pixels which is the largest box size required to have
 at least 8 boxes.
Using a maximum likelihood method one can determine
which are the best-fitting parameter values of the simulations 
(signal + noise) to the {\it COBE}--DMR data.\\
	
In Fig. 3 we show a contour plot of the likelihood obtained for 
the {\it COBE}--DMR data.
The maximum is at $Q_{{\rmn rms}-{\rmn PS}}=10^{+3}_{-2.5}  \ \mu$K and
$n=1.8^{+0.35}_{-0.65}$ (95\% marginalised errors) and the
contour level at 68\% is compatible with
the assumed standard value $Q_{{\rmn rms}-{\rmn PS}}=18  \pm 3\ \mu$K for
the Einstein-de Sitter model with a scale invariant primordial 
spectrum of density perturbations, $n=1$.
The various analysis of the 4 years {\it COBE} data combined 
give as the best-fitting parameters $Q_{{\rmn
rms}-{\rmn PS}}=15.3^{+3.8}_{-2.8}
\ \mu$K and $n=1.2\pm 0.3$. 
The result presented here predicts larger values of $n$ and
smaller values of $Q_{{\rmn rms}-{\rmn PS}}$ than the result indicated
above (although 
always inside the anticorrelation law for the two parameters). 
This result is in agreement with the one found by Smoot et al. (1994),
using a similar approach. Smoot et al. (1994)
found for the best fit, 
$Q_{{\rmn rms}-{\rmn PS}} = 13.2 \pm 2.5 \ \mu$K and
$n = 1.7_{-0.6}^{+0.3}$.
A possible explanation for the discrepancy between our results and those 
obtained with the standard methods could be a bias present in the likelihood 
estimator. In the tests of our algorithm we found a 
systematic bias in the marginalized likelihood functions both for 
$Q_{\rmn rms}$ and $n$ with tipical values of $\delta n \sim + 0.2$ and 
$\delta Q_{\rmn rms} \sim - 2$ which could explain part of our discrepancy. 
The reason for this bias can be the difference between the assumed Gaussian 
form for the likelihood of the partition function in eq. (4) and the real 
non-Gaussian distribution. 
The probability distribution of the ${\cal Z}$ at each $(q,\delta)$ 
obtained from simulations is similar to a Gaussian probability distribution 
but with a longer tail for high values.  
We also think that maybe the noise can contribute to that bias.
The high order moments (large $q$) of the partition function are very
sensitive to the tails of
 the distribution of the temperature fluctuations. A low signal to noise ratio
(as is the case for the COBE-DMR data) could raise the parameter $n$ that
best fit the COBE-DMR data. We did some tests in this direction and apparently
the noise can increase the value of $n$ (and consequently can produce a 
lower value of $Q_{rms}$).


\subsection{Multifractal analysis}
We apply the formalism of section 2.2 to the simulations
and to the {\it COBE}--DMR data.
In Fig. 4 we plot $D(q)$ and $f(\alpha )$ for
the {\it COBE}--DMR and for one model ($Q_{{\rmn rms}-{\rmn PS}}= 15
\mu$K,
$n=1.2$) 
inside the $Q_{{\rmn rms}-{\rmn PS}}$--$n$ 
degeneration with its error bars.
The $D(q)$ curve has been obtained by fitting
a power--law to the partition function in the range of scales
$2 \le \delta \le 24$ pixels, following 
Eqs. (9)
and (10)
. Note that the value of $D(0)$ is not 2 as it would be expected for 
a continuous bidimensional surface. The mask slightly lowers this value.\\
By means of a Legendre transform (Eqs. (13)  
 and (14)) 
we have obtained the corresponding $f(\alpha)$ curve.
A narrow $f(\alpha)$ curve means a very
homogeneous data. If the measure associated to the data is
multifractal in nature, these curves
should be the same for all the scale ranges. We have found that
this is not the case for the {\it COBE}--DMR data. The multifractal
curves corresponding to different scale ranges do not match
each other.
CMB simulations without noise show the same behavior.
The reason for that lies in the fact that a scaling
like  that in Eq. (9) 
is not present. This can be illustrated by 
looking at the behavior of the local slopes
of $\ln Z(q,\delta)$ vs. $\ln \delta$. In Fig. 5
we show the change in the {\it reduced} slopes ($\tau(q,\delta)/(q-1)$) 
as a function of the  scale for a fixed value of the parameter $q$ 
for the {\it COBE}--DMR data.
For this plot the analysis was performed only
in the top and bottom faces of the Quad-Cube which are not affected by 
the mask.
For a multifractal measure these curves should be horizontal straight
lines. As we can appreciate in the left panel, this is not the case for
the {\it COBE}--DMR data. The result for a simulation without noise 
is shown in the right panel.
In both cases, we do not see a neat plateau for large absolute values of $q$.
However it is not clear whether the fluctuations of the local reduced slopes
are just due to numerical noise related to the resolution of the maps 
(i.e. the limited number of pixels) or, on 
the contrary, these fluctuations are intrinsic to the measure and, therefore, 
prove that the measure is not a multifractal. Although our result neither
support nor contradict this interpretation, it seems more natural to expect
fractal behaviour
in the case that one is using the
absolute value of the relative temperature fluctuations $\Delta T/T$ 
as the measure (Mollerach et al. 98). 
As shown in that paper, $\Delta T/T$ fluctuations generated by the
Sachs--Wolfe effect behave like a fractional brownian fractal.

\subsection{Gaussianity}
To test whether the {\it COBE}--DMR data are Gaussian distributed, 
we compare the $H(q)$
curve for {\it COBE}--DMR with those curves arising from 
the best-fitting CMB Gaussian models obtained in section 3.2.
In Fig. 6 we show the plots of $H(q)$ for different grid  scales. For 
each realization, th measure is 
 rescaled in order to have dispersion equal to one. This allows to have 
 a small and equal range of $q$ values for all scales.
We would like to point out the deviation of the mean value from zero 
when $q$ moves away from zero. This is due to the fact that we have a 
finite number of pixels (i.e. a cosmic variance effect).
 The predicted behavior of equation (15) is only
true when we compute the mean over infinite values (or equivalently,
solve the integral between $-\infty$ and $+\infty$). Otherwise 
$H(q)$ is not zero at large values (positive and negative) of $q$.
Fig. 7 shows the likelihood distribution of the 1000
Gaussian realizations (with noise) and the dotted line corresponds to
the COBE value.
It is clear that the {\it COBE}--DMR is perfectly compatible
with the Gaussian hypothesis.


\section{Discussion and Conclusions}
We have shown in this work the power of the partition function to
describe CMB maps taking into account the information given at
different scales and by different moments. We have also shown the
flexibility of such a function to be used in various analyses:
standard likelihood, multifractal and Gaussian. We applied these  
analyses to the 4 years {\it COBE}--DMR data.\\

Based on the likelihood function we find the best-fitting parameters 
$Q_{{\rmn rms}-{\rmn PS}} = 10_{-2.5}^{+3} \ \mu$K and
$n=1.8_{-0.65}^{+0.35}$. It is remarkable the agreement between 
our work and the one by Smoot et al. (1994). \\
The {\it COBE}--DMR data (and the simulations of scale invariant power 
spectrum) do not show a fractal behavior,
regarding the absolute temperature map. 
On the other hand, recent galaxy surveys covering large scales ($>100$ Mpc) 
do not show either a fractal behaviour (Wu, Lahav \& Rees 1998). 
Both results allow to conclude that neither the mass distribution 
(assuming a linear bias) nor the intensity of the CMB show a 
fractal behaviour on large scales.

The partition function analysis performed shows no evidence for 
non-Gaussianity in the {\it COBE}--DMR data. This is in agreement with all the 
previous analyses of the {\it COBE}--DMR data except the one 
by Ferreira et al. (1998). 
Simulations done at higher resolution have shown the power of this 
method to discriminate between Gaussian and non-Gaussian signals. That 
analysis will be presented in a future paper.

Finally, we would like to remark that the likelihood analysis based on 
the partition function is computationally intensive. 
Actually a non optimized code applied to the COBE-DMR data
takes a few days (CPU time) to run in an Alpha server 2100 5/250.
Moreover, the computation of the partition function increases with
the number of pixels $N$ as $O(N)$.
This rate should be compared with the most widely
applied method used to compress data and to estimate cosmological parameters,
 the power spectrum of the fluctuations.
The direct computation of the power spectrum goes like $O(N\log N)$ (this
behaviour is due to the FFT). Standard brute-force approaches used to estimate
the power spectrum go like $O(N^3)$. The reason for this $O(N^3)$ rate is the
matrix inversion and determinant calculation whose dimension grows as the
number of pixels.
On the contrary, in the partition function likelihood analysis the number
of bins (or matrix dimension) of the likelihood is $N_q \times N_{\delta}$,
being this number usually well below one thousand (even for high resolution
maps). The number of moments $q$ is an arbitrary parameter independent of 
$N$ and the number of scales $\delta$ increases as $\leq O(N^{1/2})$.
The process of inverting the correlation matrix is clearly reduced in the
case of the partition function. This point makes the method useful for
forthcoming large data-sets. One can therefore consider the partition 
function as an alternative way to compress large data sets.
Furthermore, for the general situation of non-Gaussian data sets,
the partition function is clearly preferable to the power spectrum
since the former contains information on several moments of the data.

\section*{Acknowledgments}
We would like to thank R. B. Barreiro for kindly providing
her program for the simulations, L. Cay\'on for help
dealing with the {\it COBE}--DMR maps and interesting discussions.
SM acknowledges CONICET for financial support and to the 
Vicerrectorado de Investigacion de la Universidad de Valencia. 
JMD thanks the DGES for a fellowship.  
This work has been financially supported by
the  Spanish DGES,  project n. PB95-1132-C02-02 and project
n. PB96-0707, and by the Spanish CICYT,  project n. ESP96-2798-E.
The {\it COBE} data sets were developed by the NASA Goddard Space
Flight Center under the guidance of the {\it COBE}  Science Working
Group and were provided by the NSSDC.



\newpage

\begin{figure}
  \vspace{2cm}
  \caption{A realization of the multiplicative
    multifractal on a grid of $256 \time 256$ pixels.
    The grey scale denote the strength of the
    multifractal measure}
\end{figure}

\begin{figure}
   \vspace{2cm}
   \caption{$D(q)$ (left) and $f(\alpha)$ (right) for the multiplicative
    multifractal. The dotted line on the right panel represents
    the theoretical $f(\alpha)$ curve for a larger range of $q$}
\end{figure}

\begin{figure}
  \vspace{2cm}
  \caption{Contour confidence levels (68\%, 95\%, 99\%) for the likelihood
           distribution. The maximum is at $Q_{{\rmn rms}} = 10
          \ \mu$K and $n = 1.8$.}
\end{figure}

\begin{figure}
   \vspace{2cm}
   \caption{$D(q)$ (bottom) and $f(\alpha)$ (top) curves for {\it COBE}--DMR
            (dotted line) and for one model with $Q_{{\rmn rms}} =
           15 \ \mu$K and $n = 1.2$.}
\end{figure}

\begin{figure}
   \vspace{2cm}
   \caption{{\it Reduced} slopes computed between two consecutive
            scales for the {\it COBE}--DMR data (left) and for
           simulated data (without noise, right). Each curve corresponds
            to a fixed value of $q$ betwen $-1e6$ and $1e6$.}
\end{figure}

\begin{figure}
\vspace{2cm}
   \caption[junk]{$H(q)$ curves for eight different scales. From left
           to right and from top to bottom. $2 \times 2$ pixels
           to $24 \times 24$ pixels. The x-axis represents the
           values of $q$.}
\end{figure}

\begin{figure}
  \vspace{2cm}
  \caption{Likelihood distribution for 1000 simulations for the
           best-fitting model in section 3.2. The value corresponding
           to the {\it COBE}--DMR data is shown as a dotted line and is in
           clear agreement with the Gaussian hypothesis.}
\end{figure}

\label{lastpage}

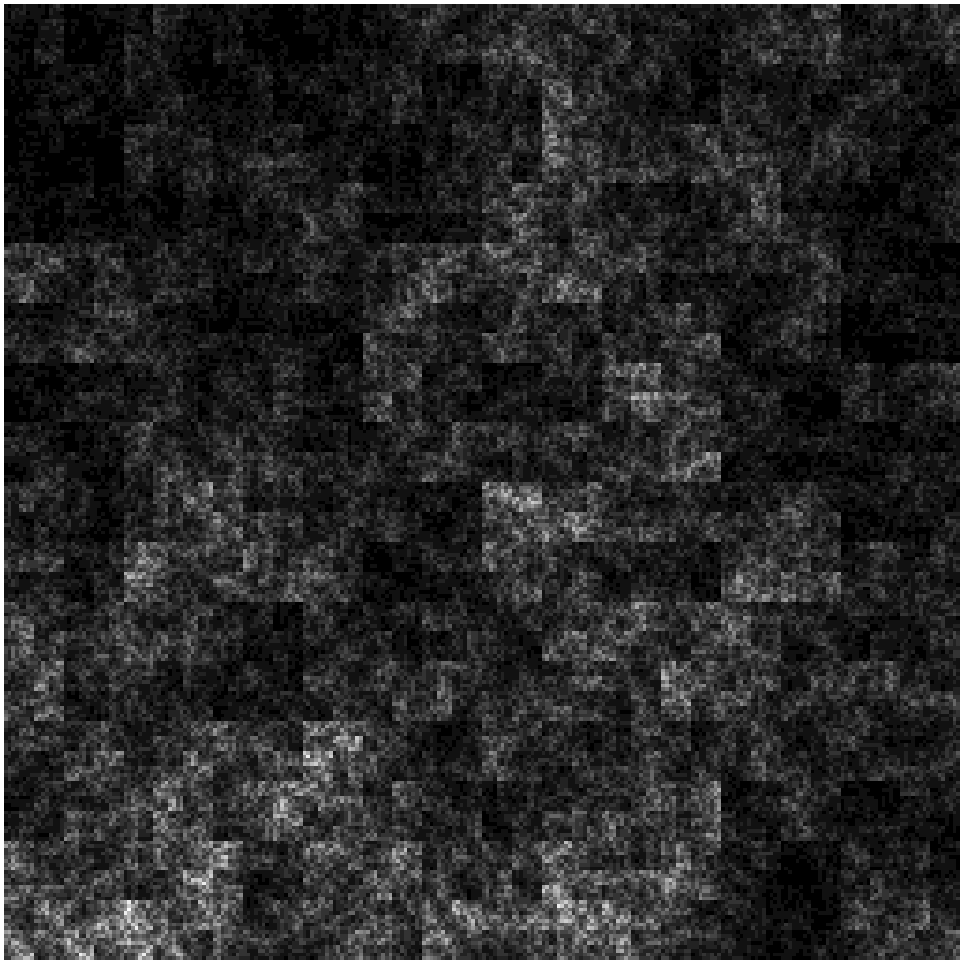
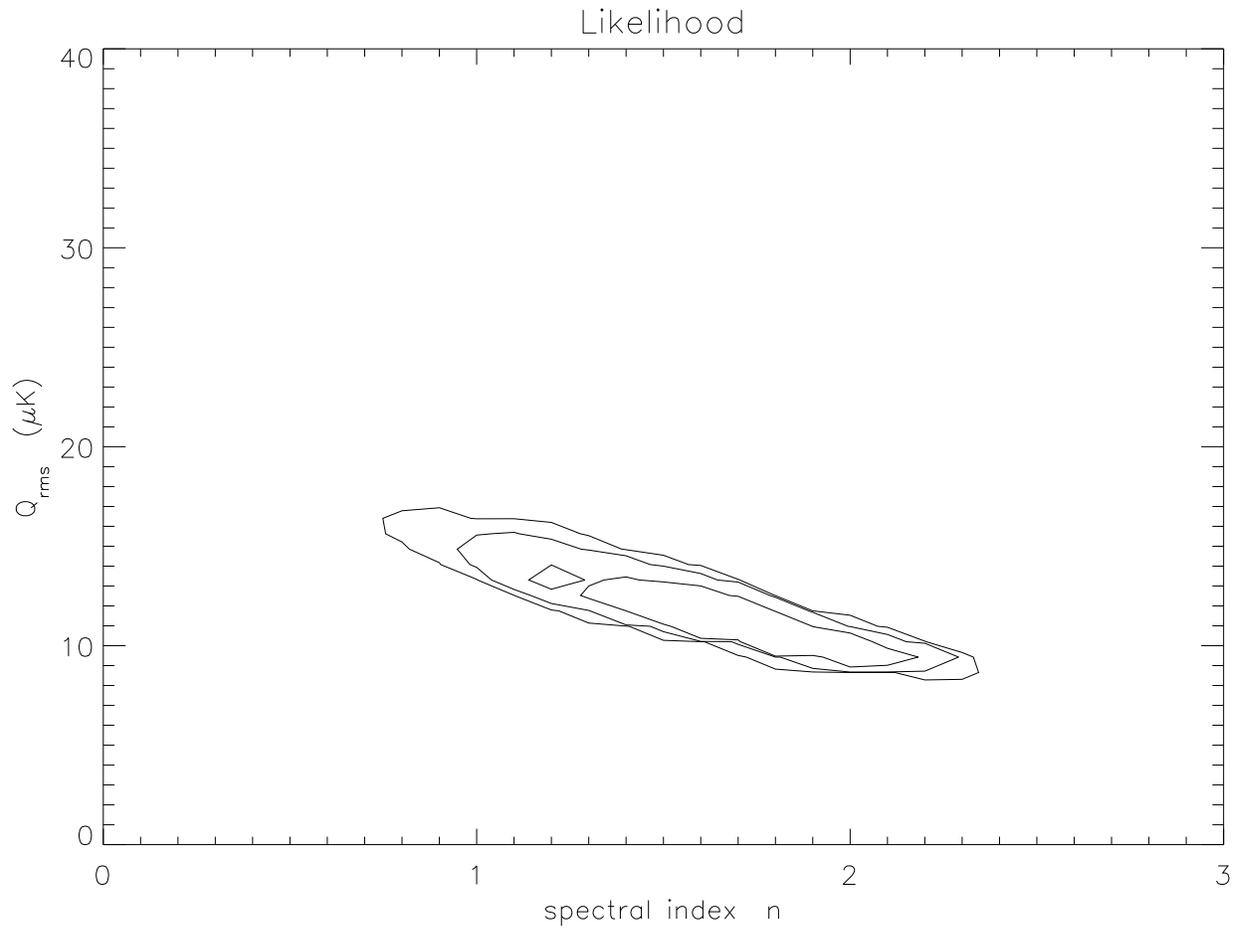
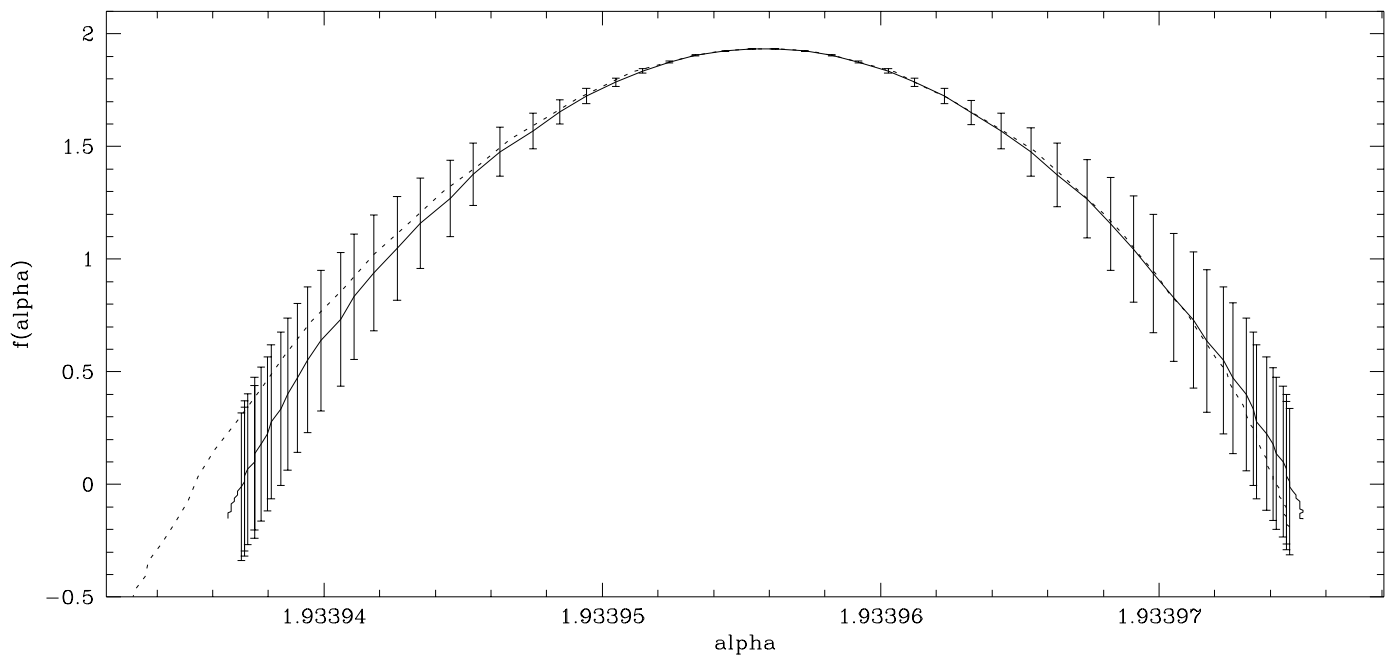
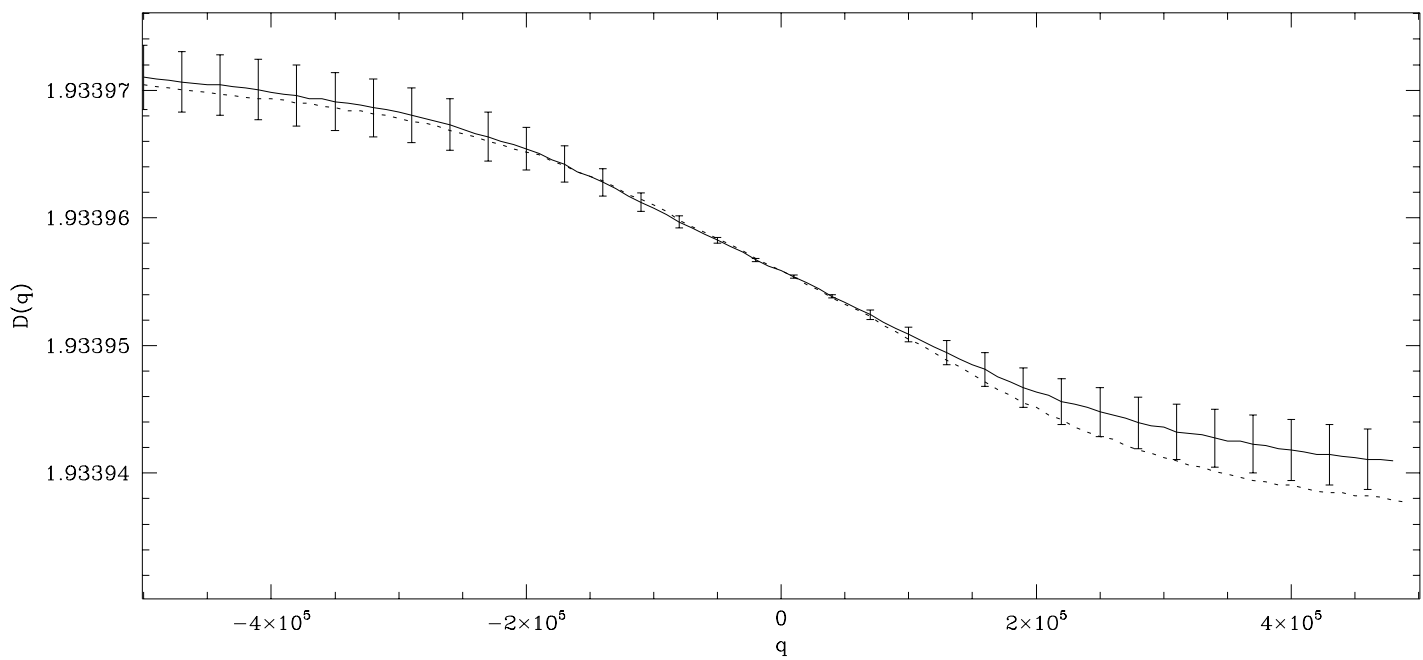
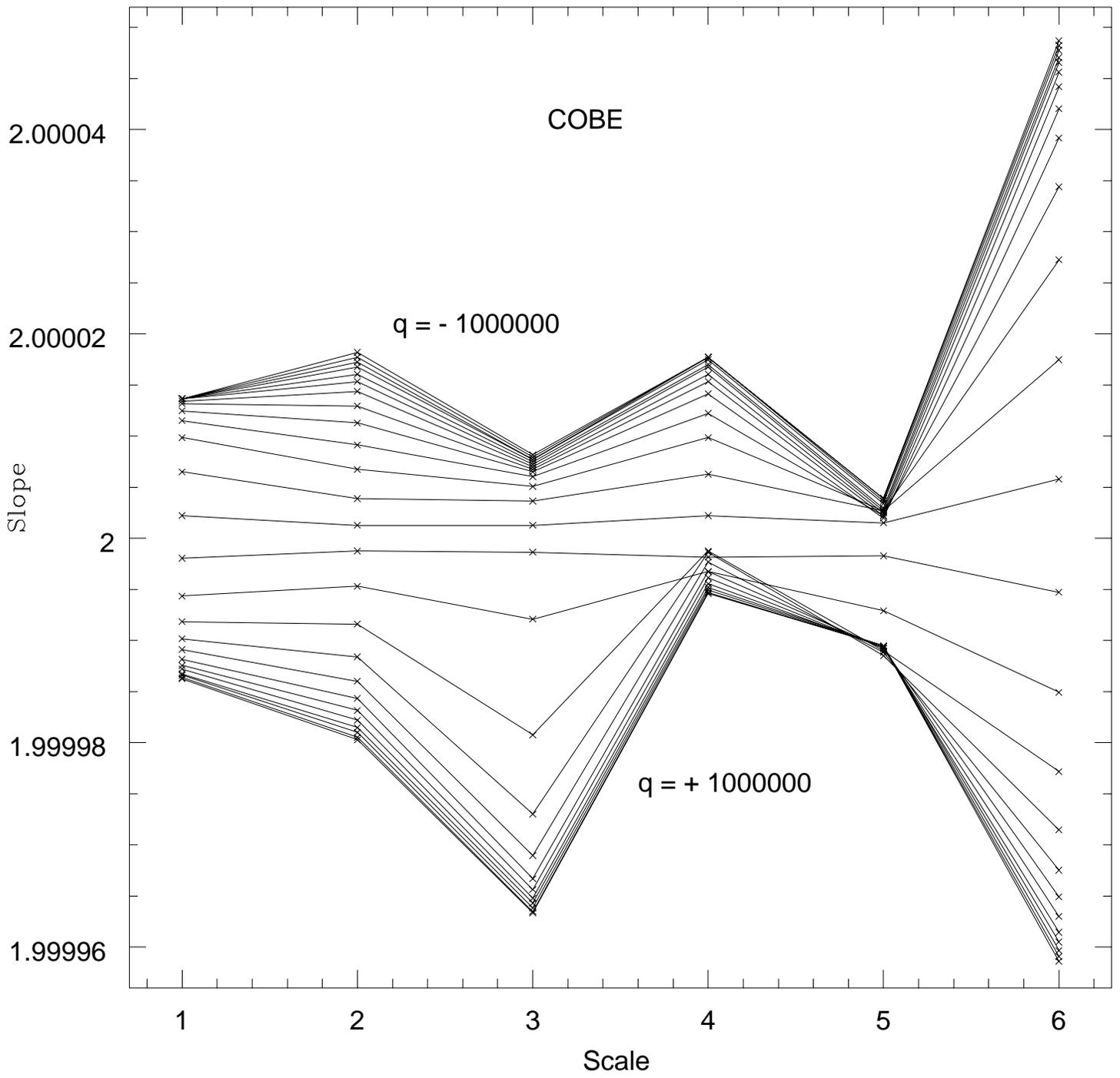
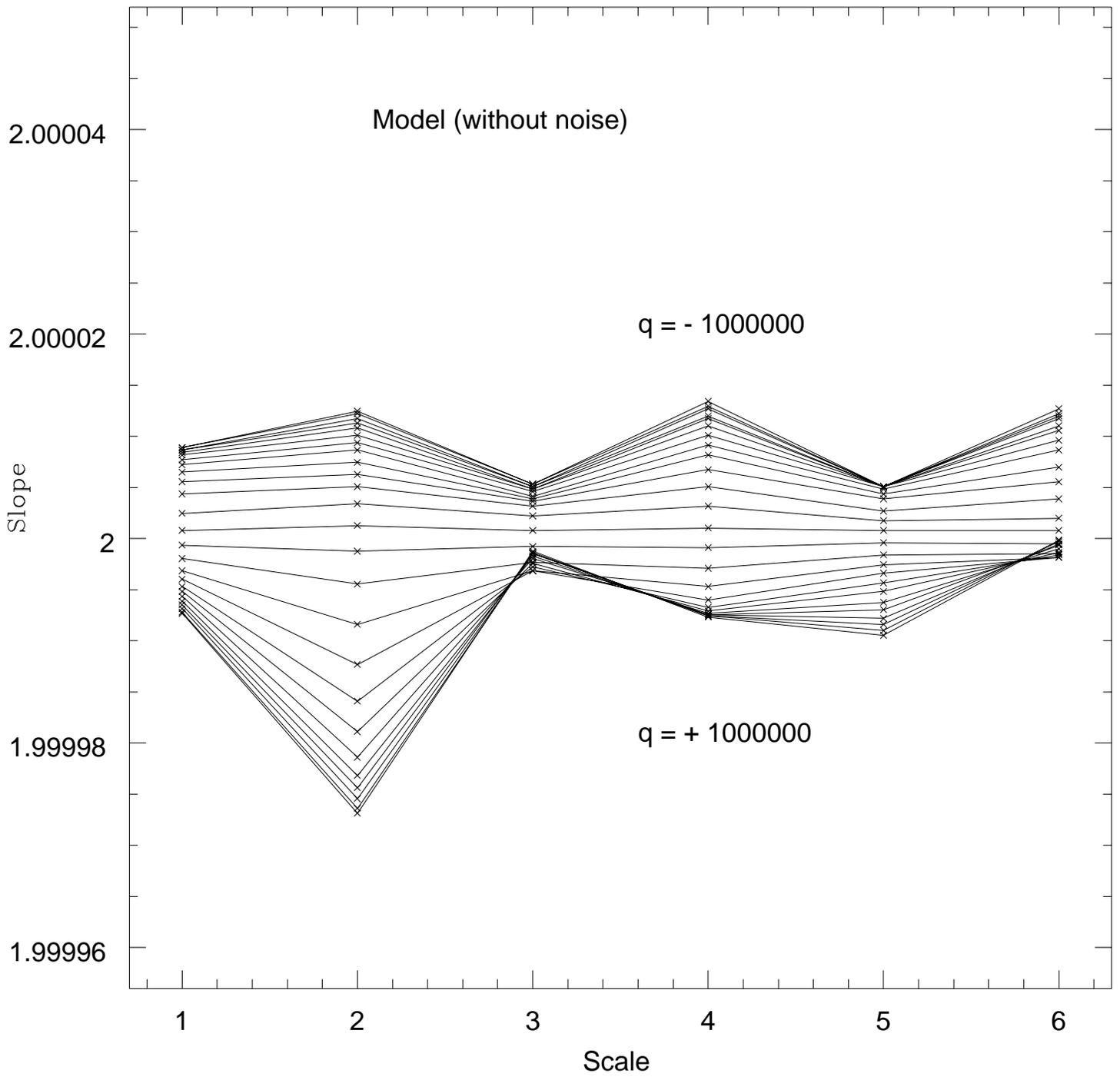
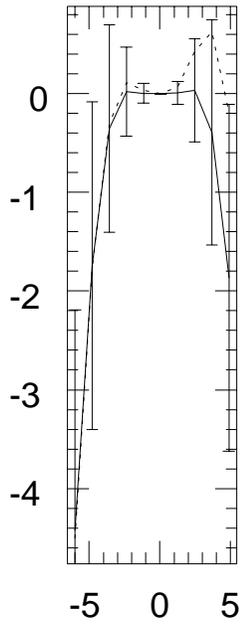
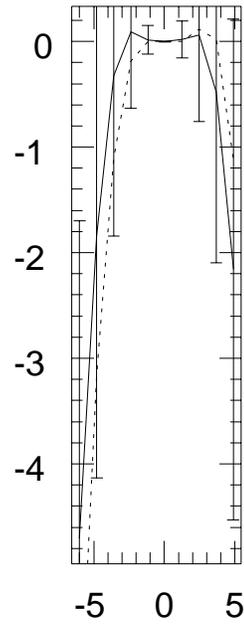
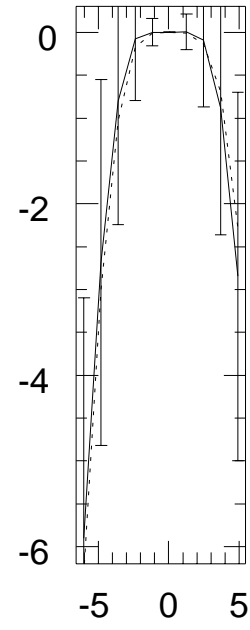
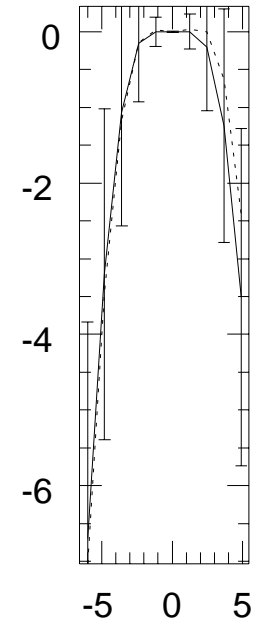
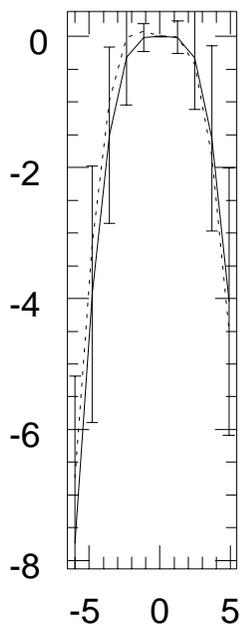
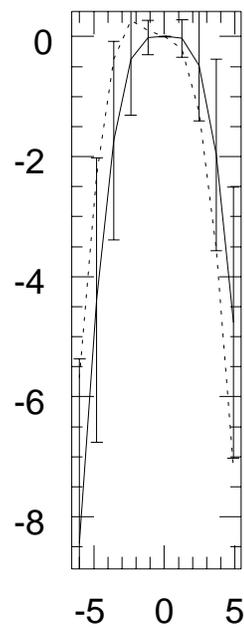
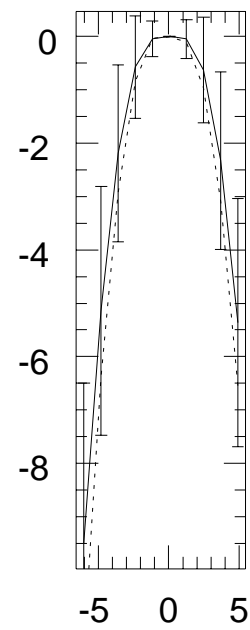
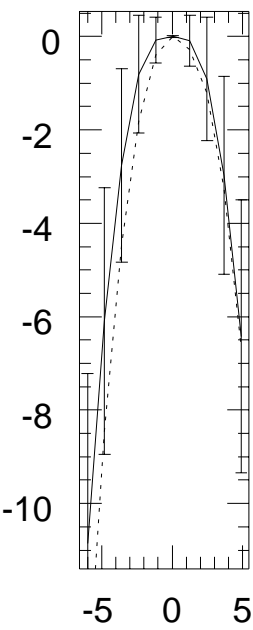
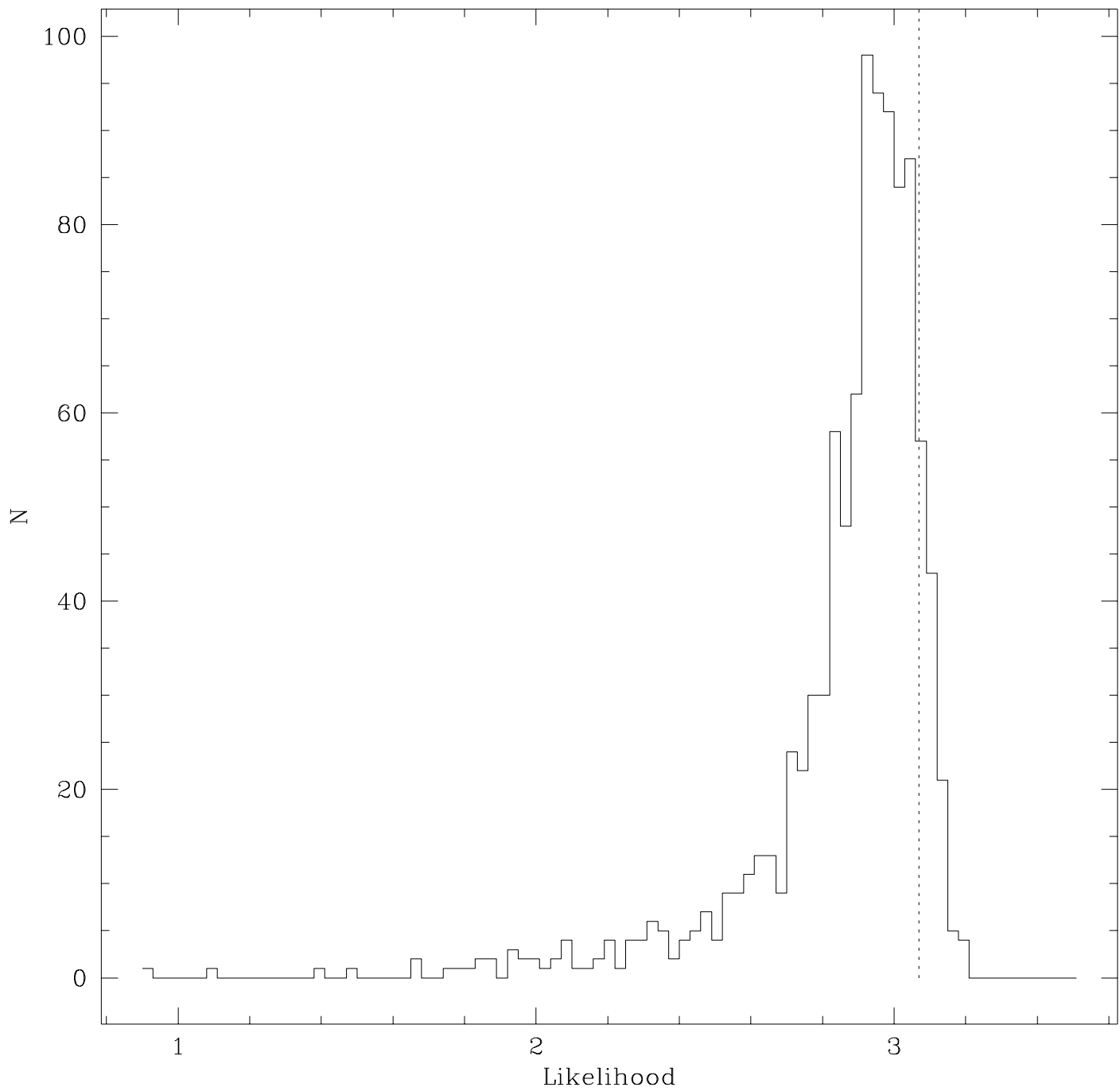
\end{document}